\documentclass{ws-p8-50x6-00}
\input{psfig}
\begin{document}
\title{Generalized Polarizabilities in a Constituent Quark Model}

\author{B. PASQUINI,$^1$ \underline{S. SCHERER},$^2$ AND D. DRECHSEL$^2$}

\address{
$^1$ ECT$^\ast$, Strada delle Tabarelle 286, 
I-38050 Villazzano (Trento), {\it and} INFN, Trento, Italy}
\address{$^2$ Institut f\"{u}r Kernphysik, Johannes Gutenberg-Universit\"{a}t,
J.~J.~Becher-Weg 45,\\
D-55099 Mainz, Germany}
\maketitle

\abstracts{ We discuss low-energy virtual Compton scattering off the proton
within the framework of a nonrelativistic constituent quark model.
   A simple interpretation of the spin-averaged generalized polarizabilities
is given in terms of the induced electric polarization (and magnetization). 
   Explicit predictions for the generalized polarizabilities obtained
from a multipole expansion are presented for the Isgur-Karl model and are 
compared with results of various models.
}   

\section{Introduction}\label{sec:intro}
   In recent years, low-energy virtual Compton scattering (VCS) off the proton
as tested in, e.g., the reaction $e^-p\to e^-p\gamma$, has attracted 
considerable interest from both the experimental\cite{Roche_2000} 
and theoretical\cite{Scherer_1999} side.   
  In fact, real Compton scattering (RCS) has a long history of providing
important theoretical and experimental tests for models of hadron structure.
   For example, the famous low-energy theorem (LET) of Low
and Gell-Mann and Goldberger\cite{LET} 
predicts the scattering amplitude for a spin-$\frac{1}{2}$ system
in terms of the charge, mass, and magnetic moment to the first two 
orders in the photon energy. 
   In principle, any model respecting the symmetries entering the derivation
of the LET should reproduce the constraints of the LET.
   It is only terms of second order which contain new information on the 
structure of the nucleon specific to Compton scattering.
   For a general target, these effects can be parametrized in terms of 
two constants, the electric and magnetic polarizabilities
$\bar{\alpha}$ and $\bar{\beta}$, respectively\cite{Klein_1955}.
   A generalized low-energy theorem (GLET) including virtual photons
was given in Refs. \cite{Guichon_1995,Scherer_1996}.
   The use of virtual photons considerably increases the possibilities to
investigate the structure of the target, because on the one hand
the energy and three-momentum of the virtual photon can be varied 
independently, and on the other hand also longitudinal components of
current operators become accessible.
   In  Ref.\ \cite{Guichon_1995}, the model-dependent response beyond the
GLET was analyzed by means of a multipole expansion.
   Only terms contributing to first order in the energy of the
outgoing real photon were kept, and the result was expressed in
terms of ten generalized polarizabilities (GPs) which are functions of
the three-momentum of the initial-state virtual photon.
   If charge conjugation and particle crossing are symmetries of the
underlying model, only six of the ten GPs are 
independent\cite{Drechsel_1997}.
   A somewhat different generalization in terms of generalized
dipole polarizabilities was obtained in Ref.~\cite{Lvov_2001},
where a covariant starting point was chosen.

\section{Compton Tensor and Low-Energy Theorem}

\label{sec:LET}
   In a covariant framework, the nucleon Compton tensor is defined as    
\begin{eqnarray}
\label{definition:vcstensor}
\lefteqn{(2\pi)^4\delta^4(p_f+q'-p_i-q) M^{\mu\nu}_{fi}(p_f,q';p_i,q)=}
\nonumber\\
&&\int d^4x\, d^4y\, e^{-iq\cdot x}e^{iq'\cdot y}
\langle N(p_f)|T[J^\mu(y)J^\nu(x)]|N(p_i)\rangle,
\end{eqnarray}
   where $T$ refers to the covariant time-ordered product. 
   As a special case, the RCS amplitude ($q^2=q'^2=0$) is obtained by 
contracting Eq.\ (\ref{definition:vcstensor}) with the polarization vectors 
$\epsilon_\nu$ and $\epsilon'^\ast_\mu$ of the initial and final photons, 
respectively,\footnote{We use natural units $\hbar=c=1$,
$e>0$, $e^2/(4\pi)\approx 1/137$.}
\begin{equation}
\label{mrcs}
{\cal M}_{\rm RCS}
=-ie^2\epsilon'^\ast_\mu\epsilon_\nu M^{\mu\nu}_{fi}(p_f,q';p_i,q).
\end{equation}
  The invariant amplitude for 
$\gamma^\ast N\to \gamma N$ reads
\begin{equation}
\label{mvcs}
{\cal M}_{\rm VCS}=
ie^2\left(\vec{\epsilon}_T\cdot\vec{M}_T
+\frac{q^2}{\omega^2}\epsilon_z M_z\right),
\end{equation}
where $\epsilon_\nu=e\bar{u}\gamma_\nu u/q^2$ is the polarization vector
of the virtual photon, and where we made use of current conservation.
   In the center-of-mass (c.m.) frame, using the Coulomb gauge for the final
(real) photon, the transverse and longitudinal parts of ${\cal M}_{\rm VCS}$ 
can be expressed in terms of eight and four independent structures, 
respectively,
\begin{equation}
\vec{\epsilon}_T\cdot \vec{M}_T=\vec{\epsilon}\,'^\ast \cdot 
\vec{\epsilon}_T A_1 + \cdots,\quad
M_z=\vec{\epsilon}\,'^\ast \cdot \hat{q} A_9 + \cdots,
\end{equation}
   where the functions $A_i$ depend on the three kinematical variables
$|\vec{q}\,|$, $|\vec{q}\,'|$, and $z=\hat{q}\cdot\hat{q}\,'$.

   Model-independent predictions for the functions $A_i$, based on
Lorentz invariance, gauge invariance, crossing symmetry, and the discrete 
symmetries were obtained in Ref. \cite{Scherer_1996}.
   For example, the result for $A_1$ up to second order in the momenta
$|\vec{q}\,|$ and $|\vec{q}\,'|$ reads 
\begin{eqnarray}
\label{a1}
A_1&=&-\frac{1}{M}+\frac{z}{M^2}|\vec{q}\,|
-\left(\frac{1}{8M^3}+\frac{r^2_E}{6M}-\frac{\kappa}{4M^3}
-\frac{4\pi\bar{\alpha}}{e^2}\right)|\vec{q}\,'|^2\nonumber\\
&&+\left(\frac{1}{8M^3}+\frac{r^2_E}{6M}-\frac{z^2}{M^3}
+\frac{(1+\kappa)\kappa}{4M^3}\right)|\vec{q}\,|^2.
\end{eqnarray}
   To this order, all functions $A_i$ are completely specified in terms of 
quantities which can be obtained from elastic electron-proton scattering and 
RCS, namely  $M$, $\kappa$, $G_E$, $G_M$, $r^2_E$, $\bar{\alpha}$, 
and $\bar{\beta}$.

\section{Generalized Polarizabilities and Their Interpretation}
\label{sec:GPs}
   A systematic analysis of the structure-dependent terms specific to VCS
off the nucleon was performed by Guichon {\em et al.}\cite{Guichon_1995}.
   The nonpole terms were expressed in terms of three spin-independent
and seven spin-dependent GPs which were introduced
in the framework of a multipole analysis, restricted to 
first order in $|\vec{q}\,'|$, but for arbitrary $|\vec{q}\,|$.

\begin{figure}[ht]
\begin{center}
\epsfig{file=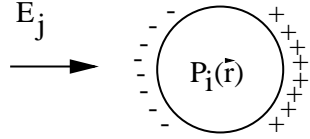,width=4cm}
\end{center}
\caption{\label{fig:electricpolarization} The electric field induces an
electric polarization.}
\end{figure}

   For a spin-0 system such as the pion or for the spin-averaged nucleon
amplitude the GPs can, to a certain extent, be interpreted in terms
of simple classical concepts\cite{Lvov_2001}. 
   To that end, let us consider a spherical charge distribution which is 
exposed to a uniform static electric field.
   The connection between the electric field and the induced
electric polarization is, to first order, given by the polarizability
tensor $\alpha_{ij}(\vec{r}\,)$ (see Fig.\ \ref{fig:electricpolarization}), 
\begin{equation}
\label{electricpolarization}
P_i(\vec{r}\,)=4\pi \alpha_{ij}(\vec{r}\,) E_j.
\end{equation}

\noindent   
   Defining
\begin{equation}
\label{alphaijr}
\alpha_{ij}(\vec{r}\,) = \int \frac{d^3 q}{(2\pi)^3}
\exp(-i\vec{q}\cdot \vec{r}\,) \alpha_{ij}(\vec{q}\,),
\end{equation}
spherical symmetry imposes the general form
\begin{equation}
\label{alphaijq}
\alpha_{ij}(\vec{q}\,)=\alpha_L(q)\underbrace{\hat{q}_i\hat{q}_j}_{
\mbox{
longitudinal}}
+\alpha_T(q)
\underbrace{(\delta_{ij}-\hat{q}_i\hat{q}_j)}_{\mbox{transverse}},
\end{equation}
where $q=|\vec{q}\,|$.
   We now consider the form factor 
\begin{equation}
\label{ffeff}
F(\vec{q}\,)=\int d^3 r \exp(i\vec{q}\cdot\vec{r}\,) 
\rho_{\rm eff}(\vec{r}\,)
\end{equation}
associated with the effective charge density 
\begin{equation}
\label{rhoeffective}
\rho_{\rm eff}(\vec{r}\,)
=\rho_0(r)-\vec{\nabla}\cdot\vec{P}(\vec{r}\,).
\end{equation}
   Inserting Eq.\ (\ref{rhoeffective}) into Eq.\ (\ref{ffeff}),
performing a partial integration, and finally applying the definition
of Eq.\ (\ref{alphaijr}) with the specific form of Eq.\ (\ref{alphaijq}),
one obtains
\begin{eqnarray}
\label{ffeffr}
F(\vec{q}\,)&=&F_0(q)-\int d^3 r\exp(i\vec{q}\cdot\vec{r}\,)
\vec{\nabla}\cdot\vec{P}(\vec{r}\,)
\nonumber\\&=& 
F_0(q)+i\vec{q}\cdot\int d^3 r \exp(i\vec{q}\cdot\vec{r}\,)
\vec{P}(\vec{r}\,)\nonumber\\
&=&F_0(q)+4\pi i q_i \int d^3 r \exp(i\vec{q}\cdot\vec{r}\,)
\alpha_{ij}(\vec{r}\,) E_j
\nonumber\\&=&
F_0(q) + 4\pi i\vec{q}\cdot\vec{E} \alpha_L(q).
\end{eqnarray}
   Equation (\ref{ffeffr}) nicely displays the connection between the
longitudinal generalized polarizability $\alpha_L(q)$ and the
modification in the charge distribution.
   The general connection between the polarizability tensor 
$\alpha_{ij}(\vec{r}\,)$ and the generalized electric polarizabilities
$\alpha_L(q)$ and $\alpha_T(q)$ of Eq.\ (\ref{alphaijq}) 
reads\cite{Lvov_2001}
\begin{eqnarray}
\label{alphaijgen}
\alpha_{ij}(\vec{r}\,)&=& 
     \alpha_L(r) \hat r_i \hat r_j
      + \alpha_T(r) (\delta_{ij} - \hat r_i \hat r_j)\nonumber\\
&&    + \frac{3\hat r_i \hat r_j - \delta_{ij}}{r^3}
        \int_r^\infty [\alpha_L(r')-\alpha_T(r')]\,r'^2\,dr'.
\end{eqnarray}
   In particular, both generalized electric polarizabilities 
are required to fully describe the induced electric polarization.
   The transverse generalized electric polarizability $\alpha_T(q)$
is associated with rotational displacements of charges which do not
contribute to $\delta\rho(\vec{r}\,)=-\vec{\nabla}\cdot\vec{P}(\vec{r}\,)$. 
   It is {\em not} contained in the approximation of Guichon 
{\em et al.}\cite{Guichon_1995}.

   The above classical considerations can easily be extended to 
nonrelativistic quantum mechanics.
     For example, for a simple harmonic oscillator model of the type
\begin{equation}
\label{harmonicoscillator}
(M\to\infty,e) \bullet------\bullet(m,-e):
\quad V=\frac{1}{2}m\omega_0^2 r^2,
\end{equation}
the form factor without external field reads
\begin{equation}
\label{ffho0}
F_0(q)=e\left[1-\exp\left(-\frac{q^2}{4\omega_0 m}\right)\right].
\end{equation}
   It is straightforward to obtain an analytical expression for the 
form factor in the presence of a static uniform electric field $\vec{E}$,
\begin{eqnarray}
\label{fqho}
F(\vec{q}\,)&=&e\left[1-\exp\left(-\frac{q^2}{4\omega_0 m}\right)
\exp\left(-ie\frac{\vec{q}\cdot\vec{E}}{m\omega_0^2}\right)\right]\nonumber\\
&=& F_0(q)+4\pi i\vec{q}\cdot\vec{E}\underbrace{\frac{e^2}{4\pi m\omega_0^2}
\exp\left(-\frac{q^2}{4\omega_0 m}\right)}_{
\mbox{$\alpha_L(q)$}}+\cdots.
\end{eqnarray} 
   As $q\to 0$, $\alpha_L(q)$ reduces to the ordinary electric polarizability
$\alpha=e^2/(4\pi m\omega^2_0)$ which is commonly interpreted as 
a measure of the stiffness or rigidity of a system\cite{Holstein_1990}.

\section{Compton Tensor in a Nonrelativistic Framework}
\label{sec:ctnf}
   Before presenting the results for the GPs in the framework of 
Guichon {\em et al.}\cite{Guichon_1995}, 
we provide a short outline of the ingredients
entering the calculation of the Compton tensor\cite{Pasquini_2001}.
   The starting point is the nonrelativistic interaction
Hamiltonian in the Schr\"{o}dinger picture,
\begin{eqnarray}
\label{hi}
H_I&=&H_{I,\,1}+H_{I,\,2},\\
\label{hi1}
H_{I,\,1}&=&\int d^3 x\,  J^\mu(\vec x) A_\mu(\vec x),\\
\label{hi2}
H_{I,\,2}&=&\frac{1}{2}
\int d^3x
\int d^3 x'
B^{\mu\nu}(\vec x,\vec x\,')A_\mu(\vec x)
A_\nu(\vec x\,'),
\end{eqnarray}
   where $A^\mu(\vec{x})$ is the second-quantized photon field and
\begin{eqnarray}
\label{vecj}
\vec J(\vec x)&=&
\sum_{\alpha=1}^N
\frac{ e_\alpha}{2m_\alpha }\delta^3(\vec x-\vec r_\alpha)
\left(\frac{\stackrel{\rightarrow}\nabla_\alpha}{i}
-\vec \sigma_\alpha\times\stackrel{\rightarrow}\nabla_\alpha\right)
+h.c.,\\
\label{rho}
\rho(\vec x)&=&\sum_{\alpha=1}^{N}e_\alpha \delta^3(\vec x-\vec r_\alpha),\\
\label{bmu0etc}
B^{\mu 0}&=&B^{0\nu}=0,\quad
B^{ij}(\vec x,\vec x\,')=\delta_{ij}\sum_{\alpha=1}^N
\frac{e_\alpha^2}{m_\alpha}\delta^3(\vec x-\vec r_\alpha)
\delta^3(\vec x\,'-\vec r_\alpha).
\end{eqnarray}
   The Compton tensor is obtained by calculating the 
contributions of $H_{I,2}$ and $H_{I,1}$ in first-order and
second-order perturbation theory, respectively,
\begin{equation}
\label{mmunu}
M^{\mu\nu}_{fi}(q', q,\vec{p}\,)=
S^{\mu\nu}_{fi}(q', q)
+T^{\mu\nu}_{fi}( q', q,\vec{p}\,),
\end{equation}
where $\vec{p}=(\vec{p}_i+\vec{p}_f)/2$.
   The contribution from the ``seagull'' terms is given by
\begin{equation}
\label{seagull}
S^{\mu0}_{fi}
=S^{0\nu}_{fi}=0,\quad
S^{ij}_{fi}=
\delta_{ij}
         \langle 0' |
         \sum_{\alpha}\frac{e^2_\alpha}{m_\alpha} 
         e^{i(\vec q-\vec q\,')\cdot\vec r\,'_\alpha}
         |0 \rangle,
\end{equation}
   where $\vec r\,'_\alpha $ refers to the intrinsic coordinates of 
particle $\alpha$ relative to the center of mass.
   Note that $S^{\mu\nu}_{fi}$ is symmetric under photon crossing,
$q^\mu\leftrightarrow -q'^\mu$ and $\nu\leftrightarrow \mu$.
   The explicit form of the direct and crossed channels reads
\begin{eqnarray}
\label{tmunuexplicit}
T^{\mu\nu}_{fi}
& = &\sum_{X}\frac{
\langle 0' |J^{\mu}(-\vec{q}\,',2\vec{p}_f+\vec{q}\,')|X \rangle
\langle X|J^{\nu}(\vec{q},2\vec{p}_i+\vec{q}\,)|0 \rangle}{
E_f(\vec{p}_f)+\omega'-E_X(\vec{p}_f+\vec{q}\,')}\nonumber\\
&&+\mbox{crossed-channel sum},
\end{eqnarray}  
   where every single intermediate state $X$ gives rise to a contribution
which is explicitly symmetric under photon crossing.
   This will be of some importance in checking the predictions for the GPs.
   For further evaluation, the current operator is divided
into intrinsic and c.m.\ contributions\cite{Friar_75,Arenhoevel_85}
\begin{equation}
\vec J(\vec q,\vec p\,)=
\vec j^{\,in}(\vec q\,)+\frac{\vec p}{M}\rho(\vec q\,),\quad
J_0(\vec q,\vec p\,)=\rho(\vec{q}\,).
\end{equation}

\section{Generalized Polarizabilities in a Constituent Quark Model}
\label{sec:gpcqm}
   Neither $S^{\mu\nu}_{fi}$ nor $T^{\mu\nu}_{fi}$ of 
Eqs.\ (\ref{seagull}) and (\ref{tmunuexplicit}), respectively,
is separately gauge invariant.
   Thus, before calculating the GPs we divide $M^{\mu\nu}_{fi}$ into two
separately gauge-invariant contributions. 
   The first consists of the ground-state propagation in
the direct and crossed channels together with an appropriately
chosen term to satisfy gauge invariance.
   The residual part contains the relevant structure information
from which one obtains the GPs.

   For the calculation of the GPs we make use of two different
expansion schemes.
   The first consists of a conventional nonrelativistic $1/M$ expansion.
   The advantage of this approach is that, at leading order in $1/M$, 
the result does not depend on the average nucleon momentum 
$\vec{p}=(\vec{p}_i+\vec{p}_f)/2$.
   In the c.m. frame $\vec{p}=-(\vec{q}+\vec{q}\,')/2$ such that 
naive photon crossing---implying the replacement of {\em any} 
$\vec{q}\leftrightarrow-\vec{q}\,'$---would turn $\vec{p}$ into $-\vec{p}$.
   However, at leading order in $1/M$, naive photon crossing is
the same as true photon crossing which, at that order, leads to constraints 
for the GPs as $|\vec{q}\,|\to 0$ (see Ref.\ \cite{Pasquini_2001} for
details).

   Here we present the results of a second scheme used by Liu {\em et al.}
in Ref.~\cite{Liu_1996}.
   There, the crossed-channel contribution is divided into a leading
plus a recoil part which is defined via matrix elements depending on 
both $\vec{q}$ and $\vec{q}\,'$.
   Furthermore, in order to incorporate, to some extent, effects due to
relativity, relativistic expressions for the energies of the intermediate 
states were used.
   Since particle crossing is not a symmetry of the nonrelativistic 
constituent quark model (NRCQM), the results
do {\em not} satisfy the relations among the ten GPs which were found
in Refs.\ \cite{Drechsel_1997}.

   The calculation is performed in the framework of the model of
Isgur and Karl \cite{Isgur_1978}, where the
$\Delta(1232)$ resonance and
low-lying negative-parity baryons 
D$_{13}$(1520), S$_{11}$(1535), S$_{31}$(1620),
S$_{11}$(1650), S$_{13}$(1700), D$_{33}$(1700)
have been included.
   In this model gauge invariance is violated first because of a mismatch 
between the resonance masses of the intermediate
states and the corresponding wave functions and second because of a 
truncation of the model space.
   On the other hand, photon crossing symmetry is respected, because
for each diagram the crossed diagram is taken into account.

\section{Results}\label{sec:results}

\begin{figure}[ht]
\centerline{\psfig{file=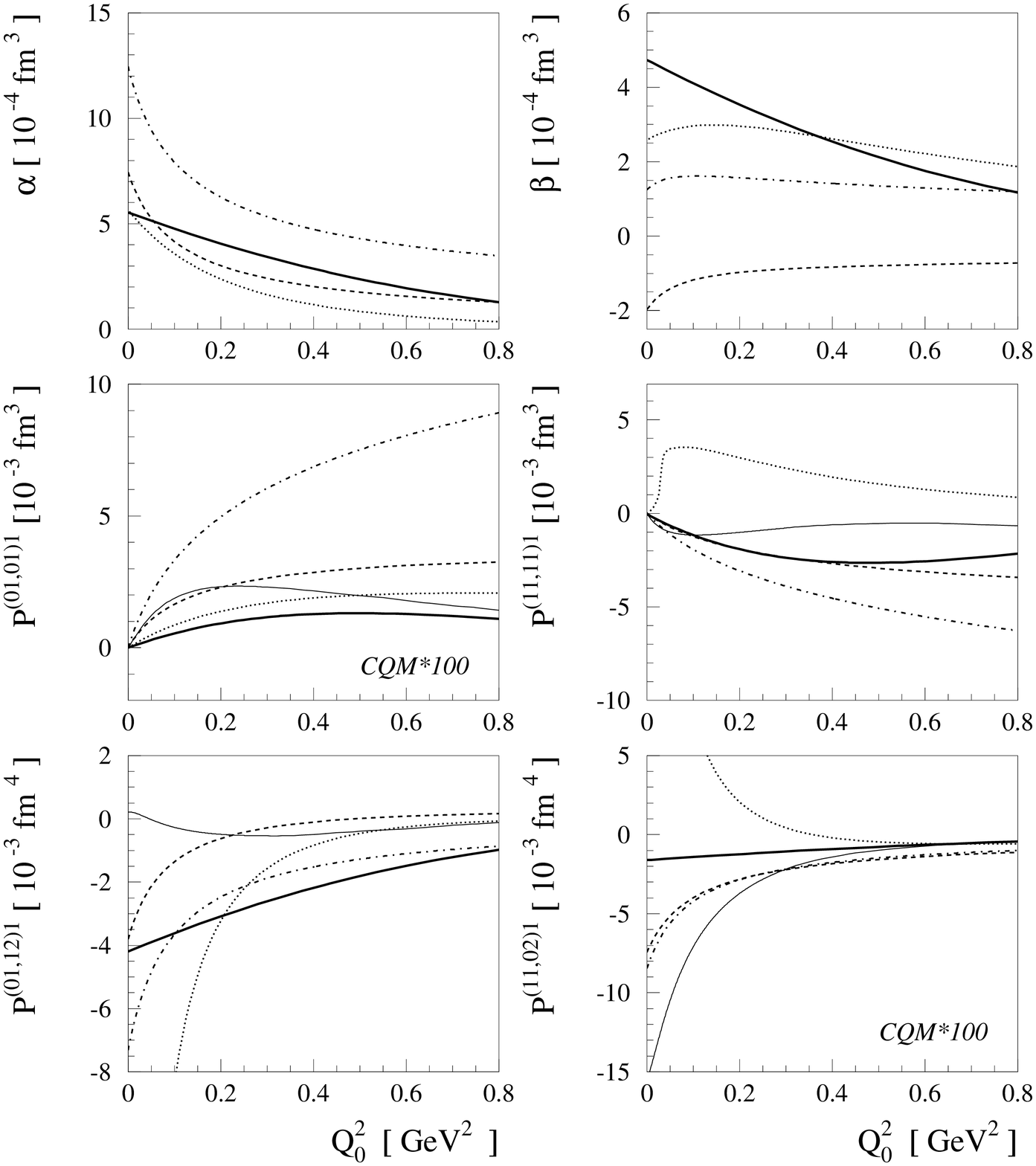,width=10cm}}
\caption{\label{comparison1}
Results for GPs in 
different model calculations as function of squared 
momentum transfer at $\omega'=0$, $Q^2_0=Q^2|_{\omega'=0}$.
Full thick lines: 
our results in the NRCQM with the scheme of Ref.\ \protect\cite{Liu_1996};
dashed lines: linear sigma model\protect\cite{Metz_1996};
dashed-dotted lines:  heavy-baryon chiral perturbation 
theory\protect\cite{Hemmert_1997};
dotted lines: effective Lagrangian model\protect\cite{Korchin_1998};
full thin lines: dispersion-relation calculation\protect\cite{Pas01}. 
Note the scaling of the CQM value for two spin polarizabilities
by a factor 100.}
\end{figure}

   The numerical results for six of the GPs are shown in 
Fig.\ \ref{comparison1}
together 
with the predictions of the linear sigma model\cite{Metz_1996} (dashed lines),
heavy-baryon chiral perturbation theory\cite{Hemmert_1997}
(dashed-dotted lines), 
an effective Lagrangian model\cite{Korchin_1998} (dotted lines), 
and dispersion relations\cite{Pas01} (full thin lines).
   For $|\vec{q}\,|\to 0$, the GPs $\alpha$ and $\beta$ reduce to the
ordinary RCS polarizabilities $\bar{\alpha}$ and $\bar{\beta}$.
   In the Isgur-Karl model, these polarizabilities
 receive their main contributions 
from the resonances D$_{13}$(1520) and  $\Delta(1232)$, respectively.
   In case of the spin GPs $P^{(01,01)1}$ and $P^{(11,02)1}$,
the NRCQM generates much smaller values than the linear $\sigma$ 
model\cite{Metz_1996} and chiral perturbation theory\cite{Hemmert_1997}.
   This is interpreted as a consequence of missing pionic degrees of freedom.

    For the generalized electric polarizability 
$$\alpha(|\vec{q}\,|)=-\frac{e^2}{4\pi}\sqrt{\frac{3}{2}} 
P^{(01,01)0}(|\vec{q}\,|),
$$
our result is in agreement with Ref.\ \cite{Liu_1996}.
   Discrepancies in $P^{(01,01)1}$, $P^{(11,11)1}$, and  
$P^{(01,12)1}$ were found to originate from the contributions of the crossed 
channel relative to the direct one (for details, see 
Ref.\ \cite{Pasquini_2001}).

   The results for those structure functions that can be accessed in
an unpolarized experiment, are shown in Table \ref{table1}.
   The values of $|\vec{q}\,|$ correspond to the RCS limit,
MIT-Bates\cite{Shaw_1997} and MAMI\cite{Roche_2000} kinematics, respectively.
\begin{table}[ht]
\caption{\label{table1}
Structure functions $P_{LL}$, $P_{TT}$, and $P_{LT}$
in $\mbox{GeV}^{-2}$.}
\begin{center}
{\begin{tabular}{|r|c|c|c|}
\hline
&$P_{LL}$&$P_{TT}$&{$P_{LT}$}\\ 
\hline
$|\vec{q}\,|=0$ MeV
&37.0
&$-0.1$ 
&$-11.2$
\\ \hline
$|\vec{q}\,|=240$ MeV
&28.7
&$-2.8$
&$-8.8$
\\ \hline
$|\vec{q}\,|=600$ MeV
&9.9
&$-5.8$
&$-3.2$
\\
\hline
\end{tabular}
}\end{center}
\end{table}
   Table \ref{table2} contains a comparison of our results with 
experiment\cite{Roche_2000}.
\begin{table}[ht]
\caption{\label{table2}
Structure functions $P_{LL}-P_{TT}/\epsilon$ and $P_{LT}$
for $|\vec{q}\,|=600$ MeV in GeV$^{-2}$ ($\epsilon=0.62$).}
\begin{center}
{\begin{tabular}{|c|c|c|}
\hline
&$P_{LL}-P_{TT}/\epsilon$&
$P_{LT}$\\ \hline
Our calculation 
&$19.2$
&$-3.2$
\\ \hline
Experiment\cite{Roche_2000}
&$23.7$
&$-5.0$\\
&
$\pm2.2\pm0.6\pm 4.3$
&$\pm 0.8\pm 1.1 \pm 1.4$\\
\hline
\end{tabular}
}\end{center}
\end{table}

\section{Summary}
   Low-energy virtual Compton scattering opens the door to a new and rich
realm of issues regarding the structure and dynamics of composite 
systems---namely, what is the {\em local} deformation of a system exposed to 
a (soft) external stimulus.
   In an expansion in terms of the energy of the final-state real photon
and the three-momentum of the initial-state virtual photon, the GLET 
specifies the accuracy needed to be sensitive to higher-order
terms in order to distinguish between different models.
   This structure information beyond the GLET can be parametrized in
terms of generalized polarizabilities.
   To some extent, these quantities have their analogies in classical physics
and can be interpreted in terms of the induced electric polarization
and magnetization.
   We discussed the virtual Compton scattering tensor in the
framework of nonrelativistic QM and made use of the model of Isgur
and Karl to obtain predictions for the GPs of the proton.
   The model does not provide the relations among the
GPs due to nucleon crossing in combination with charge 
conjugation.      
   Clearly, the calculation has its limitations regarding relativity, gauge 
invariance and chiral symmetry, leaving room for improvement in any of 
the above-mentioned directions.
   Despite these shortcomings, the predictions provide an order-of-magnitude 
estimate for the nucleon resonance contributions and as such 
are complementary to the results of the linear sigma
model and chiral perturbation theory that emphasize pionic
degrees of freedom and chiral symmetry.

\section*{Acknowledgments}
   S.\ Scherer would like to thank A.\ I.\ L'vov for numerous interesting
and fruitful discussions.
   This work was supported by the Deutsche Forschungsgemeinschaft (SFB 443).


\begin{thebibliography}{99}
\bibitem{Roche_2000}
   J.~Roche {\em et al.}, Phys.~Rev.~Lett.~{\bf 85}, 708 (2000).
\bibitem{Scherer_1999}
   See, e.g., S.~Scherer, Czech J.~Phys.~{\bf 49}, 1307 (1999), and
   references therein.
\bibitem{LET}
   F.~E.~Low, Phys.~Rev.~{\bf 96}, 1428 (1954); 
   M.~Gell-Mann and M.~L.~Goldberger, {\em ibid.}, 1433 (1954).
\bibitem{Klein_1955}
   A.~Klein, Phys.~Rev.~{\bf 99}, 998 (1955).
\bibitem{Guichon_1995}
   P.~A.~M.~Guichon, G.~Q.~Liu, and A.~W.~Thomas,
   Nucl.~Phys.~{\bf A591}, 606 (1995).
\bibitem{Scherer_1996}
   S.~Scherer, A.~Yu.~Korchin, and J.~H.~Koch, Phys.~Rev.~C {\bf 54},
   904 (1996).
\bibitem{Drechsel_1997}
   D.~Drechsel, G.~Kn\"{o}chlein, A.~Metz, and S.~Scherer,
   Phys.~Rev.~C {\bf 55}, 424 (1997);
   D.~Drechsel, G.~Kn\"{o}chlein, A.~Yu.~Korchin, A.~Metz, and S.~Scherer,
   {\em ibid. } {\bf 57}, 941 (1998); 
   {\em ibid.} {\bf 58}, 1751 (1998).
\bibitem{Lvov_2001}
   A.~I.~L'vov, S.~Scherer, B.~Pasquini, C.~Unkmeir, and D.~Drechsel,
   hep-ph/0103172, to appear in Phys.~Rev.~C {\bf 64}, (2001). 
\bibitem{Holstein_1990}
   B.~R.~Holstein, Comments Nucl.~Part.~Phys.~{\bf 19}, 221 (1990).
\bibitem{Pasquini_2001}
   B.~Pasquini, S.~Scherer, and D.~Drechsel, Phys.~Rev.~C {\bf 63}, 
   025205 (2001). 
\bibitem{Friar_75}
   J.~L.~Friar, Ann.~Phys.~(N.Y.) {\bf 95}, 170 (1975).
\bibitem{Arenhoevel_85}
   H.~Arenh\"{o}vel, NATO Advanced Study Institute
on New Vistas in Electro-Nuclear Physics, 1985, Banff, Alberta
(Plenum Press, New York, 1986).
\bibitem{Liu_1996}
   G.~Q.~Liu , A.~W.~Thomas, and P.~A.~M.~Guichon, 
   Aust.~J.~Phys.~{\bf 49}, 905 (1996).
\bibitem{Isgur_1978}
   N.~Isgur  and G.~Karl, Phys.~Rev.~D {\bf 18}, 4187 (1978).
\bibitem{Metz_1996}
   A.~Metz and D.~Drechsel, Z.~Phys. A {\bf 356}, 351 (1996);
   {\em ibid.} {\bf 359}, 165 (1997).
\bibitem{Hemmert_1997}
   T.~R.~Hemmert, B.~R.~Holstein, G.~Kn\"{o}chlein, and
   S.~Scherer, Phys. Rev. D {\bf 55}, 2630 (1997);
   Phys. Rev. Lett. {\bf 79}, 22 (1997);
   T.~R.~Hemmert, B.~R.~Holstein, G.~Kn\"{o}chlein, and D.~Drechsel,
   Phys. Rev. D {\bf 62}, 014013 (2000).
\bibitem{Korchin_1998}
   A.~Yu.~Korchin and O.~Scholten, Phys. Rev. C {\bf 58}, 1098 (1998).
\bibitem{Pas01}
   B.~Pasquini, D.~Drechsel, M.~Gorchtein, A.~Metz, and M.~Vanderhaeghen, 
   Phys.~Rev.~C {\bf 62}, 052201 (2000);
   hep-ph/0102335,
   submitted to Eur.~Phys.~J.
\bibitem{Shaw_1997}
   J.~Shaw {\it{et al.}}, MIT--Bates proposal No. 97-03, 1997.   
\end{thebibliography}
\end{document}